\documentclass[twocolumn]{aastex63}
\submitjournal{ApJL}
\usepackage{amsmath}
\usepackage{comment}

\graphicspath{{./}{figures/}}
\newcommand{\hess}{H.E.S.S.~}

\usepackage{lineno}

\begin{document}

\correspondingauthor{J.~Damascene~Mbarubucyeye, H.~Ashkar, S. J. Zhu,  B.~Reville, F. Schüssler }
\email{contact.hess@hess-experiment.eu}


\author{F.~Aharonian}
\affiliation{Dublin Institute for Advanced Studies, 31 Fitzwilliam Place, Dublin 2, Ireland}
\affiliation{Max-Planck-Institut f\"ur Kernphysik, P.O. Box 103980, D 69029 Heidelberg, Germany}

\author{F.~Ait~Benkhali}
\affiliation{Landessternwarte, Universit\"at Heidelberg, K\"onigstuhl, D 69117 Heidelberg, Germany}

\author{J.~Aschersleben}
\affiliation{Kapteyn Astronomical Institute, University of Groningen, Landleven 12, 9747 AD Groningen, The Netherlands}

\author[0000-0002-2153-1818]{H.~Ashkar}
\affiliation{Laboratoire Leprince-Ringuet, École Polytechnique, CNRS, Institut Polytechnique de Paris, F-91128 Palaiseau, France}

\author[0000-0002-9326-6400]{M.~Backes}
\affiliation{University of Namibia, Department of Physics, Private Bag 13301, Windhoek 10005, Namibia}
\affiliation{Centre for Space Research, North-West University, Potchefstroom 2520, South Africa}

\author{A.~Baktash}
\affiliation{Universit\"at Hamburg, Institut f\"ur Experimentalphysik, Luruper Chaussee 149, D 22761 Hamburg, Germany}

\author[0000-0002-5085-8828]{V.~Barbosa~Martins}
\affiliation{DESY, D-15738 Zeuthen, Germany}

\author[0000-0002-5797-3386]{R.~Batzofin}
\affiliation{Institut f\"ur Physik und Astronomie, Universit\"at Potsdam,  Karl-Liebknecht-Strasse 24/25, D 14476 Potsdam, Germany}

\author[0000-0002-2115-2930]{Y.~Becherini}
\affiliation{Université de Paris, CNRS, Astroparticule et Cosmologie, F-75013 Paris, France}
\affiliation{Department of Physics and Electrical Engineering, Linnaeus University,  351 95 V\"axj\"o, Sweden}

\author[0000-0002-2918-1824]{D.~Berge}
\affiliation{DESY, D-15738 Zeuthen, Germany}
\affiliation{Institut f\"ur Physik, Humboldt-Universit\"at zu Berlin, Newtonstr. 15, D 12489 Berlin, Germany}

% ...

\collaboration{0}{H.E.S.S. Collaboration }



\title{H.E.S.S. follow-up observations of GRB\,221009A} 



\begin{abstract}
GRB\,221009A is the brightest gamma-ray burst ever detected.
To probe the very-high-energy (VHE, $>$\!100~GeV) emission, the High Energy Stereoscopic System (H.E.S.S.) began observations 53 hours after the triggering event, when the brightness of the moonlight no longer precluded observations. We derive differential and integral upper limits using H.E.S.S. data from the third, fourth, and ninth nights after the initial GRB detection, after applying atmospheric corrections.  The combined observations yield an integral energy flux upper limit of $\Phi_\mathrm{UL}^{95\%} = 9.7 \times 10^{-12}~\mathrm{erg\,cm^{-2}\,s^{-1}}$ above $E_\mathrm{thr} = 650$~GeV.
The constraints derived from the H.E.S.S. observations complement the available multiwavelength data. The radio to X-ray data are consistent with synchrotron emission from a single electron population, with the peak in the SED occurring above the X-ray band. Compared to the VHE-bright GRB\,190829A, the upper limits for GRB\,221009A imply a smaller gamma-ray to X-ray flux ratio in the afterglow. Even in the absence of a detection, the H.E.S.S. upper limits thus contribute to the multiwavelength picture of GRB\,221009A, effectively ruling out an IC dominated scenario.
\end{abstract}


\section{Introduction}
\label{sec:introduction}

In the last few years, several gamma-ray bursts (GRBs) have been detected in Very-High-Energy (VHE, $>\!100$~GeV) gamma rays \citep{HESS180720B, magicgrb190114c-2019, hessgrb190829a-2021}. These explosive phenomena originate from the deaths of massive stars or the mergers of compact objects \cite[e.g.][]{Meszaros}. GRBs are observed as bright flashes of gamma rays, referred to as the prompt emission, followed by long-lived and slowly evolving multiwavelength afterglow emission (see \cite{GRBs_VHE_review} for a recent review of VHE observations of GRBs). The prompt emission is thought to come from interactions within the ultrarelativistic jet produced by the catastrophic progenitor event, though the development of accurate theoretical models of the physical mechanisms underlying the emission are challenging \citep{GRBs_prompt_review}. In contrast, the afterglow is produced by the jet's subsequent interactions with the surrounding environment, and during this time the jet is well described as a conical section of a decelerating spherical blast wave \cite[see][for a recent review of GRB theory]{GRBs_theory_review}. The GRB afterglow therefore provides a well-defined laboratory for studying particle acceleration under extreme conditions.

The primary emission mechanism of photons with energies $\ll$\,GeV in the afterglow is well established as synchrotron emission by a population of accelerated charged particles. 
Assuming a homogeneous magnetic field, synchrotron photons can in principle extend up to a maximum energy $\approx 100 \Gamma$\,MeV, where $\Gamma$ is the bulk Lorentz factor of the emission zone \citep{maxSynchrotron,hessgrb190829a-2021}. A second spectral component associated with inverse Compton scattering of either ambient or the synchrotron photons is expected at higher energies, though tension between observations and single-zone synchrotron self-Compton (SSC) models have been found in the VHE domain \citep{hessgrb190829a-2021}.
The properties of any detected VHE emission therefore has important ramifications for GRB studies; the unambiguous detection of an SSC component would set constraints on the physical properties of the emission zone, while strong deviations from the expected SSC spectrum could indicate the need for a more complex set of assumptions. The two VHE-bright GRBs with the most complete data sets so far have not provided a firm conclusion on this issue \citep{magicgrb190114c-2019,hessgrb190829a-2021}.

The recent GRB~221009A is the GRB with the brightest detected prompt emission, and its redshift of $z = 0.151$ --- corresponding to a luminosity distance of around 750~Mpc --- implies an isotropic equivalent energy release in the prompt emission $E_\mathrm{iso}$ of the order of $10^{54}$~erg \citep{GCN_redshift}, marking it as an extremely energetic GRB. The large $E_\mathrm{iso}$ and close proximity resulted in the potential detection for the first time of a GRB at energies above $10$~TeV and likely the very first detection of VHE photons during the prompt phase, by the Large High Altitude Air Shower Observatory (LHAASO)~\citep{GCN_LHAASO}. To further extend this picture and characterize the VHE emission in the afterglow, the High Energy Stereoscopic System (H.E.S.S.) observed GRB~221009A on the third night, as soon as it became possible following a period of bright moonlight. H.E.S.S. is sensitive to photons at energies above tens of GeV, and has so far detected VHE emission from two GRBs \citep{HESS180720B,hessgrb190829a-2021}, including a detection more than 50~hours after the initial detection for GRB~190829A. 

In this paper, we present the H.E.S.S. observations of GRB~221009A starting on the third night after the GRB detection. We discuss the observations themselves in Section~\ref{sec:observations}, and the analysis of both  H.E.S.S. data and multiwavelength data in Section~\ref{sec:analysis}. We find no significant emission from a source at the GRB position, and derive upper limits assuming an intrinsic $E^{-2}$ spectrum. In order to place these into context, we discuss the multiwavelength modeling in Section~\ref{sec:interpretation}, and conclude in Section~\ref{sec:discussion}. Throughout the paper we assume a flat $\Lambda$CDM cosmology with H$_0$ = 67.4~km~s$^{-1}$~Mpc$^{-1}$ and $\Omega_\mathrm{m} = 0.315$ \citep{Planck2020}.


\section{Observations}
\label{sec:observations}

\subsection{Initial detection}
The detection of the prompt emission of GRB~221009A was first reported by the \emph{Fermi} Gamma-Ray Burst Monitor (GBM), which triggered on the GRB on 2022-10-09 at 13:16:59~UTC \citep{GCN_GBMdetection}; we refer to the GBM trigger time as T$_0$. GeV emission was also reported by the \emph{Fermi} Large Area Telescope (LAT) \citep{GCN_LAT}. The GRB also triggered the Neil Gehrels \emph{Swift} Observatory once it became visible to the \emph{Swift} Burst Alert Telescope an hour later. This caused the satellite to automatically slew to the source, allowing for follow-up observations by the other instruments on \emph{Swift} such as the X-Ray Telescope (XRT), which reported a localization of right ascension = 19h 13m 03s, declination = +19$^\circ$ 48' 09" (J2000) with a positional uncertainty of 5.6" \citep{GCN_Swift}.


\subsection{H.E.S.S. observations}
\label{sec:HESSobservations}
H.E.S.S. is a system of five Imaging Atmospheric Telescopes located in the Khomas Highland of Namibia ($23^{\circ} 16' 18''$, $16^{\circ} 30' 00''$) at 1800\,m above sea level. Four 12-m telescopes (CT1-4)~\citep{Aharonian2006A}, each with a mirror area of 108~m$^2$, are placed in a square formation with 120\,m side. A fifth, 28-m telescope (CT5) with a mirror area of $612 \, $m$^2$ is placed in the center of the array~\citep{Holler2015}.

\begin{table*}
\centering
\small
\begin{tabular}{llcccc}
  \hline
    Calendar date & Interval & T$_\mathrm{Start} - \mathrm{T}_0$ [s] & T$_\mathrm{End} -\mathrm{T}_{0}$ [s]  & Average zenith angle [deg] & ATC \\  
        \hline
     October 11 2022 & Night 3 & $1.901\times 10^{5}$  & $1.920\times 10^{5}$   & 49.3 & 0.46\\
     October 11 2022\footnote{taken under moderate moonlight} & Night 3 & $1.922\times 10^{5} $ & $1.929\times 10^{5}$   & 52.7 & 0.44 \\
     October 12 2022 & Night 4 & $2.765\times 10^{5}$  & $2.782\times 10^{5}$   & 49.6 & 0.49\\
     October 12 2022 & Night 4 & $2.783\times 10^{5} $ & $2.800\times 10^{5}$   & 52.6 & 0.45\\
     October 12 2022 & Night 4 & $2.800\times 10^{5} $ & $2.818\times 10^{5}$   & 57.0 & 0.41\\
     October 17 2022 & Night 9 & $7.087\times 10^{5} $ & $7.104\times 10^{5}$   & 51.7  & 0.47\\
     October 17 2022 & Night 9 & $7.105\times 10^{5} $ & $7.122\times 10^{5}$   & 56.9  & 0.65\\
   \hline
   \hline
\end{tabular}
\caption{H.E.S.S. observations of GRB\,221009A. Column 2 denotes the number of nights after T$_0$. Columns 3 and 4 represent the run start and end time since T$_0$, in seconds, respectively. Column 5 shows the average zenith angle under which the observations were conducted and column 6 shows the Atmospheric Transparency Coefficient (ATC).}
\label{tab:HESS_OBS_GRB221019A}
\end{table*} 

On October 9 and 10, H.E.S.S. could not observe the GRB as the night-sky background was too high due to the full Moon. On October 11, H.E.S.S. started observations with its 12-m telescopes as soon as observing conditions allowed. During that night, an extended 32-minute observation run was taken in nominal conditions during dark time (when the Moon was still below the horizon) followed by a second run using settings optimised for observations under high levels of optical background light such as moonlight~\citep{tomankova_lenka_2022_7400326}. H.E.S.S. continued observing GRB\,221009A in the following nights. The observations were conducted under poor atmospheric conditions due to clouds and a higher aerosol content in the atmosphere due to the regular biomass-burning season \citep{biomassBurning}. The quality of the atmospheric conditions is quantified by the atmospheric transparency coefficient~\citep{HAHN2014} with lower values corresponding to lower transmission of Cherenkov light through the atmosphere (Table~\ref{tab:HESS_OBS_GRB221019A}). Nominally accepted values of the atmospheric transparency coefficient are above 0.8. As the transparency coefficients were lower than this during the H.E.S.S. observations of this GRB, a correction procedure has been applied (discussed in the next section). Additional datasets, including ones taken on other nights, are excluded from the analysis due to further degradation of the atmospheric conditions by the presence of clouds. Unfortunately, CT5 data are not available for this study. The data taken with CT1-4 are used. Table~\ref{tab:HESS_OBS_GRB221019A} summarizes the H.E.S.S. observations used in this analysis.


\section{Analysis and results}
\label{sec:analysis}

\subsection{H.E.S.S. analysis}
\label{sec:HESSanalysis}
The H.E.S.S. data acquired during the follow-up period are analyzed using the ImPACT reconstruction procedure \citep{Parsons2014} which uses an image-template--based maximum likelihood fit. The hadronic background events produced by cosmic rays are rejected with a multivariate analysis scheme \citep{Ohm2009}. The results are independently cross-checked with a separate analysis chain based on the Model Analysis \citep{deNaurois2009} which performs a log-likelihood comparison between the recorded shower images and semianalytically generated templates. Moreover, in order to correct for atmospheric disturbances, we apply a scheme developed to assess the impact of the enhanced aerosol levels in the atmosphere on the instrument response functions derived from Monte-Carlo simulations. The scheme calculates a correction factor to the expected Cherenkov light by comparing the actual transmission profile with the ideal one used in the simulations. The correction is then applied by modifying \emph{a posteriori} the instrument response functions and reconstructed event energies \citep{Holch_2022}. These corrections are cross-checked by an analysis that uses dedicated {\it runwise} simulations taking into consideration the actual observation conditions and telescope configuration during the GRB\,221009A H.E.S.S. observations following the method outlined in~\citet{Holler_2020}. We use \textit{loose cuts}~\citep{Aharonian2006A} for the selection of gamma-ray showers. In the high-level analysis, we converted our data into GADF format\footnote{\url{https://gamma-astro-data-formats.readthedocs.io/en/latest/index.html}}~\citep{deil_christoph_2022_7304668}, and use the open source analysis package GAMMAPY~\citep{gammapy, gammapy_zenodo} (v1.0).

\begin{figure*}[htp!]
  \centering
\includegraphics[width=\textwidth]{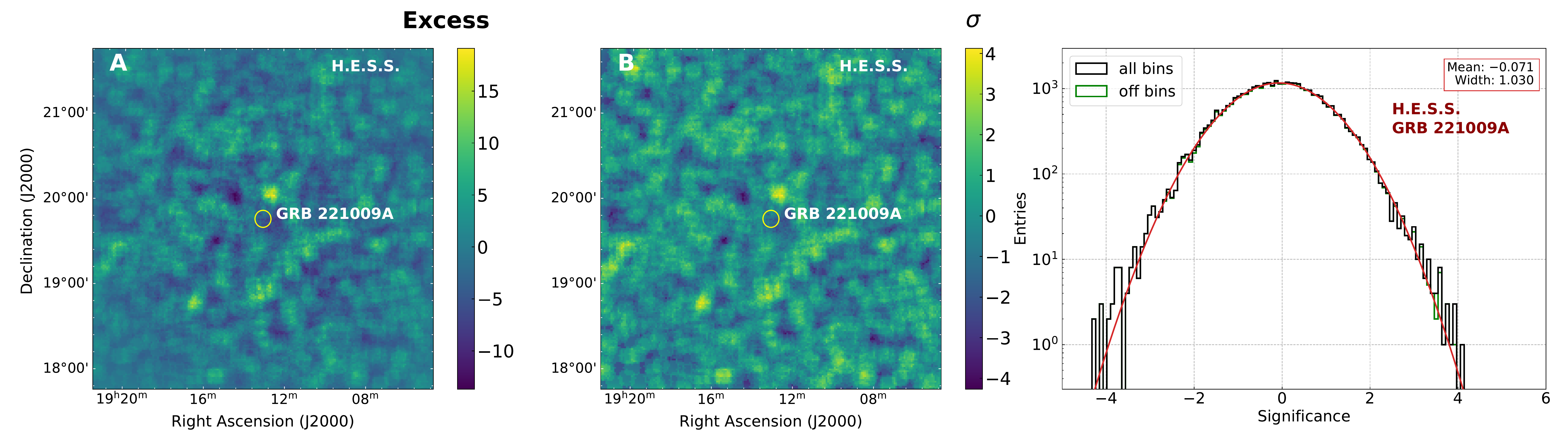}
\caption{Left: Excess count map computed from the H.E.S.S. observational data taken on GRB 221009A presented in Table~\ref{tab:HESS_OBS_GRB221019A} with a 0.1$^{\circ}$ oversampling radius (yellow circle). Middle: Significance map computed from the H.E.S.S. excess count map of GRB 221009A. Right: Significance distribution of the H.E.S.S. significance map entries in black and a Gaussian distribution fit in red.}
    \label{fig:HESS_RingMaps}
\end{figure*}

In order to search for a possible signal, and avoid accidentally including emission from other sources, we generate maps of excess gamma-ray counts and significances within a range of +/- 2.0 degrees from the expected emission position. These maps are generated using the ring background technique~\citep{RingBg} with circular ON regions of $\mathrm{0.122\,deg}$ centered at each point on the map, and corresponding annular OFF-source regions centered on the same positions with radii $\mathrm{0.5}$ to $\mathrm{0.8\,deg}$ deg to measure pure background. We exclude a circular region of $\mathrm{0.3\,deg}$ around the expected emission region from the OFF-source regions. When computing the exposure ratio between the ON and OFF-source region at each test position, a radially-symmetric model for the background acceptance within the field of view of each observation was integrated spatially over the regions. For all three nights combined, we obtain on the position of the source $\mathrm{N_{ON} = 39}$ events and $\mathrm{N_{OFF} = 686}$ events, with a ratio of on-source exposure to off-source exposure of 0.0638. Using the statistical formulation described in~\citet{LiMa} we calculate the excess counts to be $-4.8$ and we find $\mathrm{N_{ON}}$ events to be in agreement with the expected background at $-0.7\,\sigma$ level. The excess and significance maps are derived and shown in Figure~\ref{fig:HESS_RingMaps}.

\begin{figure}[htp]
  \centering
\includegraphics[width=1.1\linewidth]{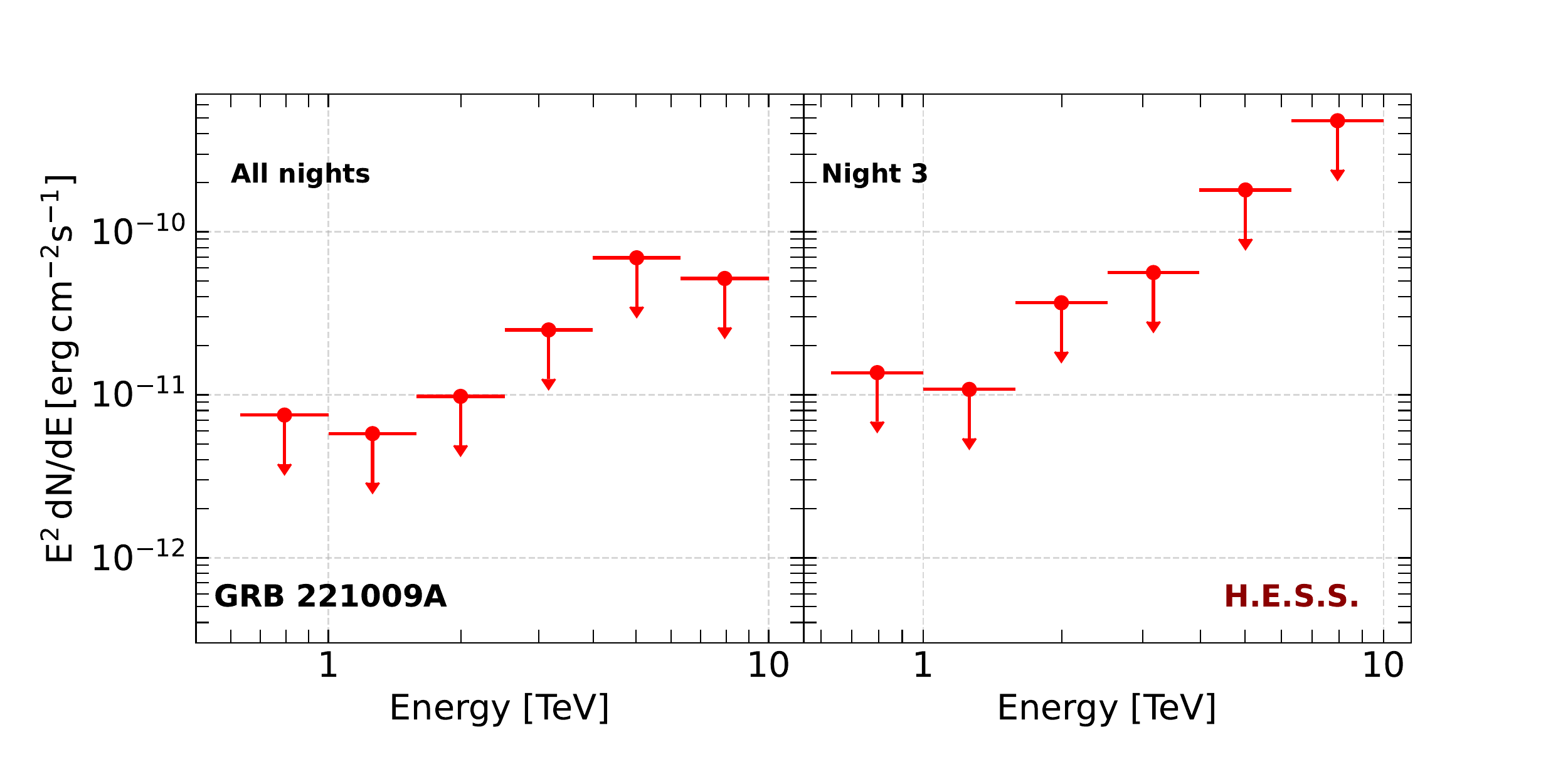}
\caption{95\% C.L. differential flux upper limits on an intrinsic (EBL-corrected) $E^{-2}$ GRB spectrum, derived from the H.E.S.S. observational data taken on GRB\,221009A in all nights combined (left) and Night 3 only (right). }
\label{fig:GRB221009A_DiffUL}
\end{figure}

Following this analysis, we detect no significant emission of VHE gamma rays at the GRB location in the combined dataset nor for each night separately. We thus compute upper limits to constrain the VHE gamma-ray emission from the GRB at the time of H.E.S.S. observations using the Reflected Background method described in \cite{RingBg} with same-size circular ON and OFF regions. The energy threshold $E_\mathrm{thr}$ sets the lower limit of the spectral analysis and is defined as the lowest energy at which the bias between reconstructed and simulated energies is below 10\%. We find a value of $E_\mathrm{thr} = 650$~GeV for the dataset combining all observations. We assume a generic intrinsic $E^{-2}$ dN/dE spectrum, and use the redshift of the source $z = 0.151$ and the model of the extragalactic background light described in~\cite{dominguez2011}, and compute 95\% confidence level (C.L.) flux upper limits using a Poisson likelihood method described in~\citet{Rolke}. The differential upper limits are shown in Figure~\ref{fig:GRB221009A_DiffUL}. Detailed results are available on a dedicated webpage\footnote{\url{https://www.mpi-hd.mpg.de/hfm/HESS/pages/publications/auxiliary/2023_GRB_221009A}}.

We compute integral flux upper limits between $E_\mathrm{thr}$ and 10~TeV for each night, where the upper bound is chosen to be the energy above which N$_\mathrm{OFF} < 10$.
The upper limits of the integral energy flux are shown in Figure~\ref{fig:GRB221009A_LC}. The combined dataset yields an integral energy flux upper limit of $\Phi_\mathrm{UL}^{95\%} = 9.7 \times 10^{-12}~\mathrm{erg\,cm^{-2}\,s^{-1}}$, and per-night integral energy flux upper limits are given in Table~\ref{tab:spectralAnalyses}. Systematic effects include uncertainties of the atmospheric corrections, the assumed intrinsic energy spectrum, differences in EBL absorption models and general uncertainties in the flux and energy scale. The systematic uncertainties are conservatively estimated to be about a factor of 2, with expected worsening of systematics with energy, for both the differential and integral upper limits on the gamma-ray flux.

\subsection{\emph{Swift}-XRT analysis}
\emph{Swift}-XRT is an X-ray imaging spectrometer with an energy range of 0.3 to 10~keV \citep{XRT_instrument}. The XRT data are obtained using the Time-Sliced Spectra tool\footnote{\url{https://www.swift.ac.uk/xrt_spectra/addspec.php?targ=01126853&origin=GRB}} \citep{Evans+2009}. The time intervals are chosen to overlap with the H.E.S.S. observations. 
There are no simultaneous XRT observations on two of the three H.E.S.S. nights, so we instead define the time ranges in such a way that they encompass one set of contiguous XRT observations immediately before and after the H.E.S.S. observations. However, using this rule for the first night of H.E.S.S. observations resulted in a too-low XRT exposure time. Hence, for this night we extend the time range to include \emph{two} sets of contiguous XRT observations on either side. (Note that, as we are using larger time bins for the XRT observations, we are underestimating what the true uncertainties would be for strictly contemporaneous observations.) In our analyses, we only include Photon Counting (PC) data. By using the data products from the Time-Sliced Spectra tool and selecting only the PC mode data, we avoid contamination from the dust rings \citep{GCN_dustRings,Evans_dustRings}.

\begin{table*}[!ht]
    \centering
    \small
    \setlength{\tabcolsep}{6pt}    
    \begin{tabular}{l || r@{ $\times$ }l  c c c | c}
      \hline
        Interval & \multicolumn{2}{c}{Time since T$_0$} & $\alpha$ & $k \times 10^{-2}$ & XRT en.\,flux $\times 10^{-11}$ & \hess en.\,flux UL $\times 10^{-11}$\\ 
         & \multicolumn{2}{c}{[s]} & & [ph keV$^{-1}$ cm$^{-2}$ s$^{-1}$] & [erg cm$^{-2}$ s$^{-1}$] & [erg cm$^{-2}$ s$^{-1}$] \\
            \hline
       Night 3 & $(1.68 - 2.22)$ & $10^{5}$ & $1.69 \pm 0.10$ & $2.14 \pm 0.30$ & \,$14.9\,\,\, \pm 2.3\,\,\,\,\,\,\,$ & 4.06\\
       Night 4 & $(2.61 - 2.90)$ & $10^{5}$ & $1.90 \pm 0.12$ & $1.31 \pm 0.20$ & $7.80 \pm 1.33$ & 1.77\\
       Night 9 & $(6.85 - 7.25)$ & $10^{5}$ & $1.85 \pm 0.25$ & $0.23 \pm 0.09$ & $1.42 \pm 0.62$ & 2.85\\
       \hline
       \hline
       \end{tabular}
       \caption{Results of analyses of XRT and H.E.S.S. data. The entries in the first column correspond to the second column of Table~\ref{tab:HESS_OBS_GRB221019A}. The second through fourth columns show the results of fitting XRT data in time intervals bracketing the nights during which H.E.S.S. observed the GRB, with 1$\sigma$ uncertainties (statistical only). The last column lists the H.E.S.S. energy flux upper limits for the time interval defined by the third and fourth columns of Table~\ref{tab:HESS_OBS_GRB221019A}. The XRT energy flux is calculated in the 0.3--10~keV range and the \hess energy flux in the 0.65--10~TeV range.}
       \label{tab:spectralAnalyses}
\end{table*}

\begin{figure}[htp]
  \centering
\includegraphics[width=\linewidth]{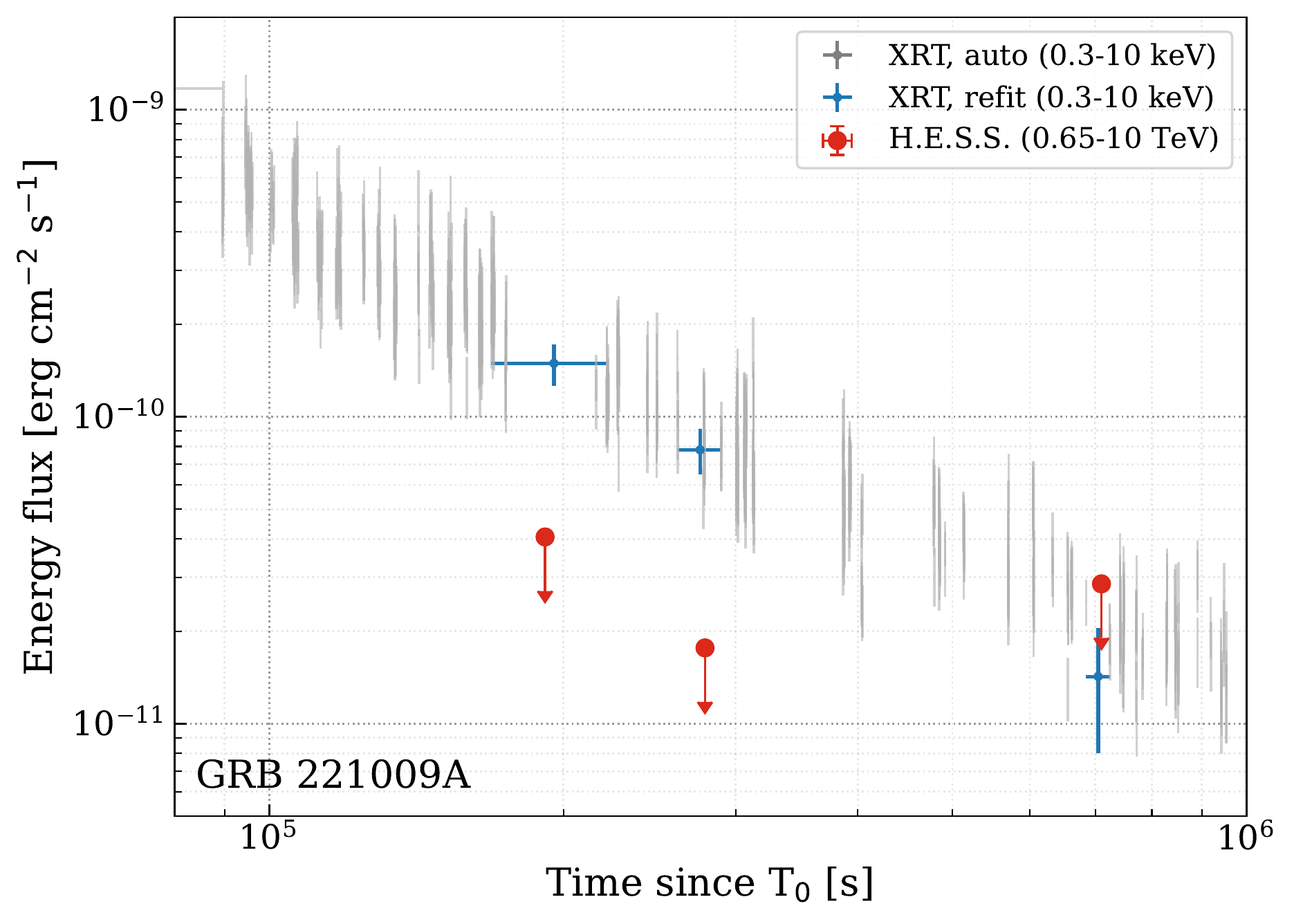}
\caption{The H.E.S.S. integral energy flux upper limits (red circles; 95\% C.L.) are derived assuming an intrinsic $E^{-2}$ spectrum. The automated XRT data (gray) are obtained from the Burst Analyser\footnote{\url{https://www.swift.ac.uk/burst_analyser/01126853/}} \citep{Evans+2010}; multiple XRT observations around the H.E.S.S. observations are then combined and refit (blue, 1$\sigma$ uncertainty). Note that the Burst Analyser assumes a larger value of intrinsic absorption than we find in our analysis and therefore returns a larger unabsorbed energy flux. The extension of the H.E.S.S. error bars in the $x$ direction, depicting the duration of the H.E.S.S. observations, is smaller than the size of the markers.}
\label{fig:GRB221009A_LC}
\end{figure}

The XRT data are fit using XSPEC v12.13.0c with a power law of the form $dN/dE = k(E/E_0)^{-\alpha}$ ($E_0 = 1$~keV) and two absorption components \citep{Evans+2009}. More specifically, we fit the data with the model \texttt{TBabs * zTBabs * powerlaw} with the Galactic column density $N_\mathrm{H,gal} = 5.38\times10^{21}$~cm$^{-2}$ \citep{galactic_absorption}. We simultaneously fit the three nights with the column density at the source $N_\mathrm{H,int}$ tied across all spectra but free to vary, under the assumption that the intrinsic absorption does not vary on these timescales. We keep $k$ and $\alpha$ free in each time interval. We define the fitting statistic to be the \texttt{C-statistic}\footnote{\url{https://heasarc.gsfc.nasa.gov/xanadu/xspec/manual/XSappendixStatistics.html}}, suitable for XRT data. Assuming a constant $N_\mathrm{H,int}$, we obtain $N_\mathrm{H,int} = (1.32\pm0.18) \times 10^{22}$ cm$^{-2}$. The results are presented in Table~\ref{tab:spectralAnalyses} and plotted in Figure~\ref{fig:GRB221009A_LC}, and are compatible with those presented in \cite{221009A_XRays}, which similarly reports a softening in the X-ray spectrum on these timescales.

The assumption of constant column density on these timescales has been challenged in other GRBs, recently by \cite{Campana+_changingAbsorption} for GRB~190114C. For GRB~221009A, indications of a higher degree of absorption at earlier times have indeed been noted in the optical data \citep{221009A_optical}. Because there is some degeneracy between the effects of $N_\mathrm{H,int}$ and $\alpha$ (e.g., a larger value of $N_\mathrm{H,int}$ can be somewhat compensated by a softer value of $\alpha$) this also has an effect on the returned best-fit photon spectrum. If $N_\mathrm{H,int}$ is indeed higher around Night 3 ($\approx$~2 days after T$_0$) than the later nights, then the true value of $\alpha$ for Night 3 should be softer than the returned 1.7 and therefore more similar to the value of 1.9 that we find for the other two H.E.S.S. nights (although we note that the indices are consistent within 2$\sigma$). 
A thorough study of this effect is beyond the scope of this paper, so for the purposes of the discussion in Section~\ref{sec:interpretation}, we do not require that our modeling explain the XRT data on Night 3 very strictly.

\subsection{Fermi-LAT analysis}
The \emph{Fermi}-LAT is a pair conversion telescope that detects gamma rays between tens of MeV and hundreds of GeV \citep{LAT_instrument}. We perform an unbinned likelihood analysis of \emph{Fermi}-LAT data over time ranges spanning each set of H.E.S.S. observations (Table~\ref{tab:HESS_OBS_GRB221019A}) using \texttt{gtBurst} v. 03-00-00p5 \citep{gtburst}. 
We use the \texttt{P8R3\_SOURCE} event class, recommended for analyses on these timescales, and the corresponding instrument response functions\footnote{\url{https://fermi.gsfc.nasa.gov/ssc/data/analysis/documentation/Cicerone/Cicerone_LAT_IRFs/IRF_overview.html}}. We select events in the energy range 100~MeV and 10~GeV within 12$^\circ$ of the burst position and a zenith angle cut of 100$^\circ$. We use the \texttt{PowerLaw2} model\footnote{\url{https://fermi.gsfc.nasa.gov/ssc/data/analysis/scitools/source_models.html}} for the GRB spectrum, and the latest Galactic (fixed normalization) and isotropic templates for the background\footnote{\url{https://fermi.gsfc.nasa.gov/ssc/data/access/lat/BackgroundModels.html}}, and include all catalog sources \citep{4FGL} within 20$^\circ$ of the GRB position. No significant emission from the GRB (Test Statistics $< 1$) is observed in the \emph{Fermi}-LAT data during the H.E.S.S. observations, so 95\% C.L. upper limits are computed assuming an $E^{-2}$ spectrum. We find differential energy flux upper limits (between 100~MeV and 10~GeV) of $7.1\times10^{-10}$ and $2.6\times10^{-10}$ erg\,cm$^{-2}$\,s$^{-1}$ during Night 3 and Night 4, respectively; the upper limit for Night 3 is shown in Figure~\ref{fig:MWL_SED}.  

\section{Discussion}
\label{sec:interpretation}

\begin{figure}[htp]
  \centering
\includegraphics[width=\linewidth]{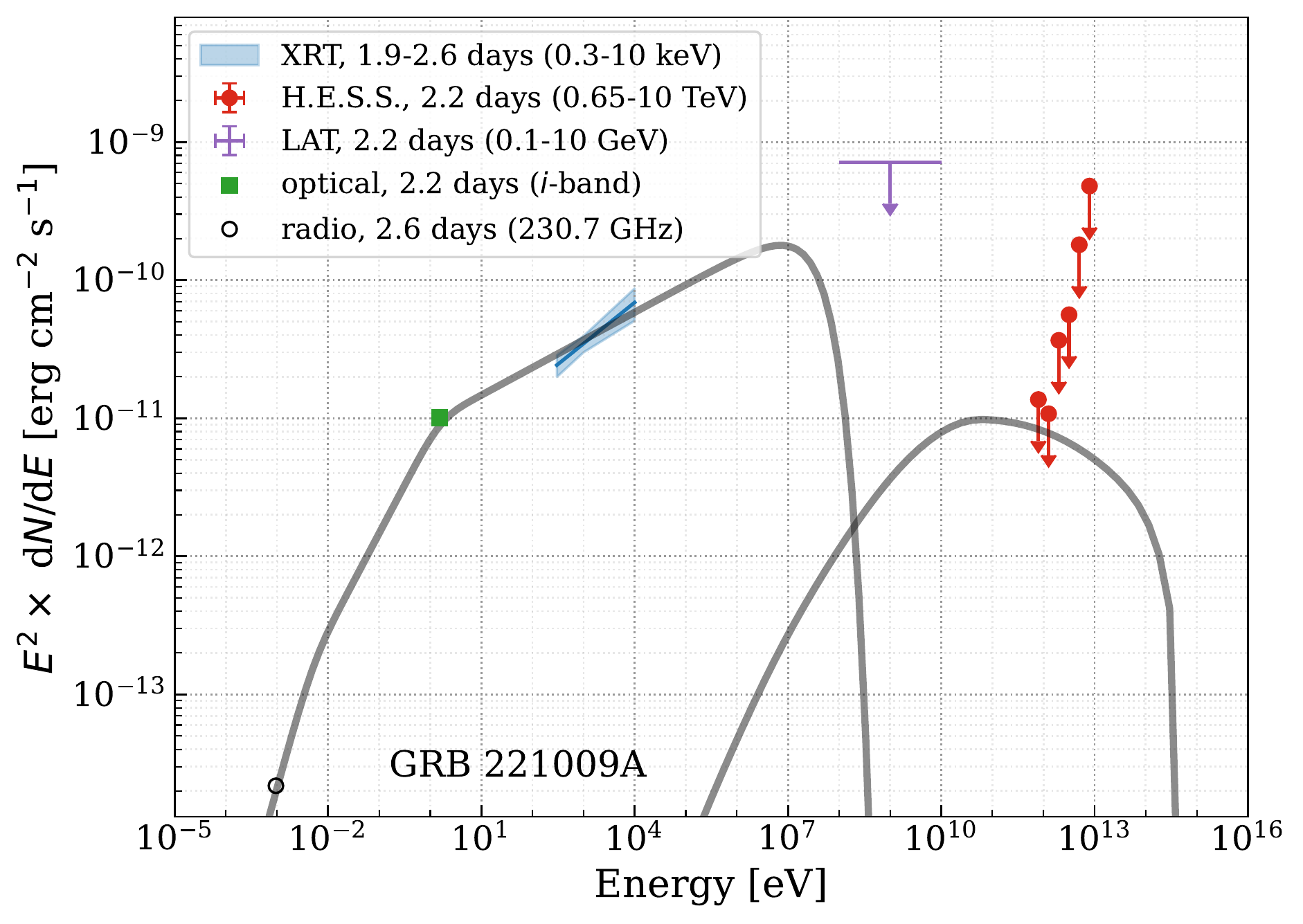}
\caption{The H.E.S.S. 95\% upper limits on Night 3 (red) are plotted along with the XRT (blue, $1\sigma$) best-fit spectrum and LAT (purple, 95\% C.L.) upper limit, as well as publicly available radio data from the Submillimeter Array (black open circle; \cite{GCN_radio}) and an optical flux (green square; extracted from Figure~2 of \cite{221009A_optical}) in a multiwavelength SED. An example set of synchrotron and SSC emission components --- arising from a single, partially cooled electron population described in Table~\ref{tab:model} --- are also shown to illustrate a possible explanation of the multiwavelength observations.}
\label{fig:MWL_SED}
\end{figure}

The \hess upper limits, when combined with multiwavelength observations, can be used to
constrain possible emission scenarios of GRB 221009A several days after the prompt event. Figure~\ref{fig:MWL_SED} shows the spectral energy distribution (SED) including the results of the H.E.S.S., XRT, and LAT analyses. Also included are the optical $i$-band flux during the time of the H.E.S.S. observations \cite[extracted from Figure 2 of][]{221009A_optical} and a publicly available radio observation \citep{GCN_radio} that most closely matches the time window of the first night of \hess observations. 

The SEDs on all nights during which \hess measurements took place are consistent with synchrotron emission from a single electron population. In such a model, the synchrotron spectrum peaks above the energy range covered by the XRT, implying  Klein Nishina (KN) suppression of any inverse Compton (IC) component. For IC-dominated cooling, KN suppression could account for the hard cooled spectrum of electrons emitting in the XRT range \cite[e.g.][]{Agaronyan,Nakar,Breuhaus}. However, this is difficult to reconcile with the \hess upper limits in the VHE range, as electrons producing X-ray photons via synchrotron emission should produce a comparable or greater gamma-ray flux via their IC emission. The energy density of optical photons was insufficient for internal absorption of TeV photons, while the LHAASO detection suggests that absorption on a local external field was unlikely. 
Ruling out IC-dominated cooling, we consider instead models in which synchrotron dominated cooling can account for the multiwavelength observations. We adopt a single-zone thin shell model \cite[see][]{Huangetal22}, assuming self-similar expansion of a relativistic shock following an impulsive point-like explosion \citep{BlandfordMcKee}. 

Within this model, we assume the radio to X-ray emission is produced by single population of continuously injected electrons. The photon index of the XRT emission (see Table \ref{tab:spectralAnalyses}) was consistent with $\alpha \approx 1.8$ on all nights \cite[see also][]{221009A_XRays}. Data from the Nuclear Spectroscopic Telescope Array (NuSTAR) indicate this spectrum continued unbroken above the XRT range \citep{2023arXiv230204388L}. These measurements imply either an uncooled electron population $dN_{\rm e}/d\bar{\gamma}_{\rm e} \propto \bar{\gamma}_{\rm e}^{-2.6}$, or the cooled spectrum assuming continuous injection: $dN_{\rm inj}/d\bar{\gamma}_{\rm e} \propto \bar{\gamma}_{\rm e}^{-1.6}$. Here $\bar{\gamma}_{\rm e}$ is the electron Lorentz factor in the fluid frame. For the cooling break to remain above the NuSTAR range (3-79\,keV), requires specific conditions \cite[see for example][eq. 14]{Huangetal22}. This scenario was considered by \citet{2023arXiv230204388L}, who modelled the synchrotron component assuming a wind profile. To match the measured flux, more than half of the downstream internal energy needs to be converted to non-thermal electrons and magnetic field energy, with these two components being in near equipartition. The SSC flux is negligible in such a scenario. To maintain the cooling break above the X-ray range requires a low mass loss rate, and though low mass loss rates are expected from the polar regions of low-metallicity stars \citep{2012A&A...546A..42M}, it is unclear if such a wind profile can be sustained over a large distance from the progenitor.  

We consider the alternative possibility of a cooled hard injection spectrum, comparing against the first night of \hess\ observations.
To match the flux levels, we introduce a parameter $\eta_{\rm inj}\leq1$, the ratio of non-thermal particle density flux, to the particle density flux downstream. The injection energy is left as a free parameter, and since the total integrated energy for an injection spectrum with index $<2$ is determined by the maximum electron energy, the is fixed by the energy efficiency. The model parameters provided in Table \ref{tab:model} were chosen to match the selected measurements, while just reaching the \hess\ upper limits. The cooling break in the synchrotron component occurs between the radio and optical data points, at a flux level comparable to the \hess\ upper limit. As synchrotron cooling dominates for the chosen parameters, the corresponding IC flux remains below the upper limits.
 
\begin{table}
\centering
\setlength{\tabcolsep}{8pt}    
     \begin{tabular}{l c}
       \hline  \hline
        Explosion energy $E$ & $10^{54}$\,erg\\
         External density $n_{\rm ext}$ & $1.7$\,cm$^{-3}$\\
         Injection fraction $\eta_{\rm inj}$ & $0.1\%$\\
         Electron equipartition fraction $\epsilon_e$ & $9 \times 10^{-4}$\\
         Magnetic equipartition fraction $\epsilon_B$ & $8\times 10^{-4}$ ($0.07$G)\\ 
        \hline
        \hline
        \end{tabular}
         \caption{Parameters used for single zone model fit, adopting the constant external-density solution of \citet{BlandfordMcKee}. We choose $\bar{\gamma}_{\rm min} = 0.66 m_{\rm p}/m_{\rm e}$ where $m_{\rm p}/m_{\rm e}$ the proton to electron mass ratio. }
         \label{tab:model}
 \end{table}  

\section{Summary and conclusion}
\label{sec:discussion}
H.E.S.S. began observing GRB\,221009A approximately 53 hours after the initial \emph{Fermi}-GBM detection. The observations were taken under less-than-optimal atmospheric conditions caused by clouds and aerosols. 
No significant VHE signal is detected on the third, fourth, and ninth nights after the detection, nor in the combined dataset of all three nights. 
When combining the data from all three nights, we find a 95\% upper limit on the 0.65--10~TeV energy flux of $\Phi_\mathrm{UL}^{95\%} = 9.7 \times 10^{-12}~\mathrm{erg\,cm^{-2}\,s^{-1}}$. 

The \hess upper limits help constrain possible emission scenarios when compared with the multiwavelength observations. The X-ray spectra on the three \hess nights were found to remain hard, with photon indices ranging from 1.7 to 1.9. Taken together with the approximately contemporaneous optical and radio data, these measurements suggest synchrotron emission from a single electron population, with continuous injection of either an uncooled soft spectrum, or a cooled hard spectrum. The photon energy spectrum peaks above the XRT range, \cite[and also that of the NuSTAR,][]{2023arXiv230204388L} and KN suppression of any inverse Compton emission is unavoidable. An IC-dominated loss scenario appears to be ruled out by the \hess upper limits. In contrast, the multiwavelength SED of the nearby low-luminosity GRB~190829A, which was detected at TeV energies three nights after the prompt emission, was better described by a single component from X-rays to VHE gamma rays. Those data were consistent with photon indices $\approx 2$ for both the XRT and \hess energy ranges on all nights \citep{hessgrb190829a-2021}. The results from GRB~221009A, potentially the brightest ever detected GRB, highlight the distinct character of these two bursts, both in terms of their non-thermal particle acceleration and emission properties. As discussed in \cite{hessgrb190829a-2021} \cite[see also][]{Huangetal22,Salafiaetal}, an accurate reproduction of the MWL observations of GRB~190829A is challenging within a single-zone SSC framework.
Other theoretical models put forward to account for the multiwavelength measurements of GRB~190829A, including external Compton \citep{Zhangetal21} or two-zone  \citep{Khangulyan} models serve to highlight the necessity for high quality spectral and temporal data of GRBs and their afterglows at all available wavelengths to understand the underlying physical mechanisms at play.  

With upper-limits in the VHE band, we consider only a single-zone model for the first night of \hess observations, assuming an electron population whose cooled spectrum is consistent with the inferred XRT spectrum. An alternative uncooled electron scenario was considered in \cite{2023arXiv230204388L} \cite[see also][]{Sato}. The hard injection scenario requires a spectral index deviating substantially from that predicted from shock acceleration theory \cite[e.g.][]{Achterberg}. Harder spectra have been predicted from other acceleration schemes such as relativistic shear acceleration \citep{Rieger} or converter mechanisms \citep{2003PhRvD..68d3003D, Stern03,2019ApJ...880L..27D}. A detailed consideration of the underlying acceleration process is however beyond the scope of the current paper, though the emerging multiwavelength dataset for GRB~221009A will provide a valuable data set for future theoretical studies. Our results highlight the role imaging atmospheric Cherenkov telescopes have in improving our understanding of these powerful transient events.

\section*{Acknowledgments}
We thank Lauren Rhodes for discussions on the radio and optical data, and Phil Evans for assistance in analyzing XRT data. This work made use of data supplied by the UK Swift Science Data Centre at the
University of Leicester.

The support of the Namibian authorities and of the University of
Namibia in facilitating the construction and operation of H.E.S.S.
is gratefully acknowledged, as is the support by the German
Ministry for Education and Research (BMBF), the Max Planck Society,
the German Research Foundation (DFG), the Helmholtz Association,
the Alexander von Humboldt Foundation, the French Ministry of
Higher Education, Research and Innovation, the Centre National de
la Recherche Scientifique (CNRS/IN2P3 and CNRS/INSU), the
Commissariat à l’énergie atomique et aux énergies alternatives
(CEA), the U.K. Science and Technology Facilities Council (STFC),
the Irish Research Council (IRC) and the Science Foundation Ireland
(SFI), the Knut and Alice Wallenberg Foundation, the Polish
Ministry of Education and Science, agreement no. 2021/WK/06, the
South African Department of Science and Technology and National
Research Foundation, the University of Namibia, the National
Commission on Research, Science \& Technology of Namibia (NCRST),
the Austrian Federal Ministry of Education, Science and Research
and the Austrian Science Fund (FWF), the Australian Research
Council (ARC), the Japan Society for the Promotion of Science, the
University of Amsterdam and the Science Committee of Armenia grant
21AG-1C085. We appreciate the excellent work of the technical
support staff in Berlin, Zeuthen, Heidelberg, Palaiseau, Paris,
Saclay, Tübingen and in Namibia in the construction and operation
of the equipment. This work benefited from services provided by the
H.E.S.S. Virtual Organisation, supported by the national resource
providers of the EGI Federation.

\software{Astropy \citep{astropy}, matplotlib \citep{matplotlib}, numpy \citep{numpy}, gammapy \citep{gammapy, gammapy_zenodo}, XSPEC \citep{xspec}, gtburst \citep{gtburst}}

\bibliographystyle{aa}
\bibliography{main}

\allauthors

\end{document}